\documentstyle[twoside,fleqn,espcrc2,epsfig]{article}

\newcommand{\AmS}{{\protect\the\textfont2
A\kern-.1667em\lower.5ex\hbox{M}\kern-.125emS}}

\title{QCD-Factorization of inclusive $B$ decays and
$|V_{ub}|$ \thanks{Talk presented at the Sixth International Conference
on Hyperons, Charm and Beauty Hadrons, IIT, Chicago, June 27--July 3 2004.} 
\hfill \normalsize \tt CLNS 04/1889 }

\author{Bj\"orn O.~Lange \address{Institute for High-Energy
Phenomenology, Newman Laboratory for High-Energy Physics, \\ 
Cornell University, Ithaca, NY 14853, U.S.A.}}

\begin{document}

\begin{abstract}
Recent progress in the theoretical description of inclusive 
$B\to X_u\,l^-\bar\nu$ decays in the shape-function region is
reported. Finite moments of the shape function are related to HQET
parameters. Event fractions for several experimental cuts are
presented, with a particular emphasis on the hadronic variable 
$P_+ = E_H - |\vec P_H|$. The aim of this talk is to introduce the
$P_+$ spectrum, to compare it to the hadronic invariant mass spectrum
and the charged-lepton energy spectrum, and to study the prospect of
evaluating $|V_{ub}|$ in the presence of a large background from 
$B\to X_c$ decays.
\end{abstract}

\maketitle

\section{INTRODUCTION}

Studies of both exclusive and inclusive semileptonic decays of the $B$
meson can be used to extract the magnitude of the CKM matrix element
$|V_{ub}|$. Both methods require some input from theory. Because the
exclusive modes suffer from large form-factor uncertainties, the
determination from inclusive $B\to X_u\,l^-\bar\nu$ decays are
theoretically favored. The overall relative uncertainty on $|V_{ub}|$
measurements is currently about 15\% \cite{Battaglia:2004ti}, and
there is hope for a significant reduction of the dominant theoretical
errors in the future. The main problem in the computation of the
relevant quantities is that it requires a framework which includes a
systematic treatment of both perturbative corrections (including
Sudakov resummation) and non-perturbative effects (shape
function). Much progress in QCD-Factorization of inclusive $B$ decays
has been made in the past several months
\cite{Bauer:2003pi,Bosch:2004th,Bosch:2004bt}, and the aforementioned
problem is now understood using a sophisticated effective field theory
machinery. The main ingredients are Soft-Collinear Effective Theory
(SCET) \cite{Bauer:2000ew,Bauer:2000yr,Bauer:2001yt}, Heavy-Quark
Effective Theory (HQET) \cite{Neubert:1993mb}, and
Renormalization-Group (RG) evolution of bilocal operators on the light
cone \cite{Lange:2003ff}.

\begin{figure}[t!]
\hspace{5mm} \epsfig{file=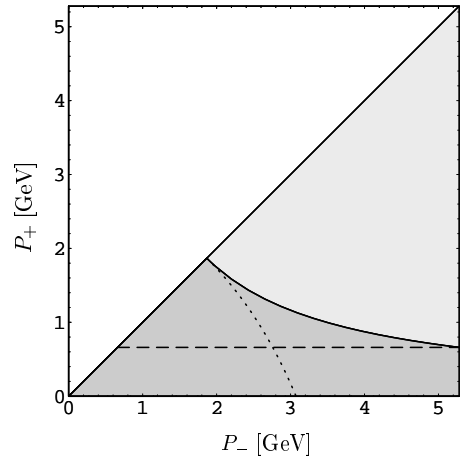, width=6cm}
\vspace{-5mm}
\caption{\label{fig:phasespace} Hadronic phase space in the light-cone
variables $P_+$ and $P_-$. The dark gray area is free of $B\to X_c$
background events, which occur in the light gray wedge above. For an
explanation of the dashed and dotted curve, see the text.}
\end{figure}

Let us begin by discussing the kinematic setup used in the description
of a typical $B\to X_u\,l^-\bar\nu$ event. The momentum of the
hadronic $X_u$ system is denoted by $P_H^\mu$, and we work in the rest
frame of the $B$ meson. It is useful to describe the hadronic phase
space in the variables $P_+=E_H-|\vec P_H|$ and $P_-=E_H+|\vec P_H|$,
in which it takes the simple form of a triangle (neglecting the pion
mass)
\begin{equation}
  0\le P_+\le P_-\le M_B\;.
\end{equation}
The phase space is depicted in Figure~\ref{fig:phasespace}. The use of
the light-cone variables $P_+$ and $P_-$ makes it easy to interpret an
event (a point) in this picture: the region of low recoil is along the
diagonal, while an event with a maximally recoiling $X_u$ is located
on the right. Events with final-state hadronic invariant masses larger
than $M_D$ are located in the light-gray wedge in the upper part of
the Figure, since $M_X^2 = P_+ P_-$.  The dark-gray region underneath
is free of the $B\to X_c$ background, which is about 60 times larger
than the $B\to X_u$ signal.

There are many possibilities to eliminate the background by cutting on
kinematic variables. Let us briefly mention a few of them. Clearly a
cut on $M_X^2 \le M_D^2$ is the ``ideal'' separator, denoted as a
solid line in Figure~\ref{fig:phasespace}. The dashed horizontal line
is located at $P_+ = M_D^2/M_B$, which marks an alternative way of
eliminating charm events. A cut on the charged-lepton energy 
$E_l \ge (M_B^2-M_D^2)/2M_B$ samples the same hadronic phase space as
a cut on $P_+ \le M_D^2/M_B$. However, it contains far fewer
events. Finally, a cut on the dilepton momentum squares 
$q^2 \le (M_B-M_D)^2$ leads to the area underneath the dotted line.

The phase space is most densely populated in the 
``shape-function region'' of large $P_- \sim O(M_B)$ and small 
$P_+ \sim O(\Lambda_{\rm QCD})$. In order to determine $|V_{ub}|$ from
this inclusive decay mode it is necessary to know what fractions of
events survive the various experimental cuts. In this talk we report
on progress to answer this question. In addition, we advertise cutting
on $P_+$ as an efficient method for future $|V_{ub}|$
determinations. To this end we study the advantages and disadvantages
of this method over comparable ones, in particular the 
``ideal separator'' $M_X$.

\section{FACTORIZATION OF THE DIFFERENTIAL DECAY RATE}
\label{sec:method}

\begin{figure}[t!]
\epsfig{file=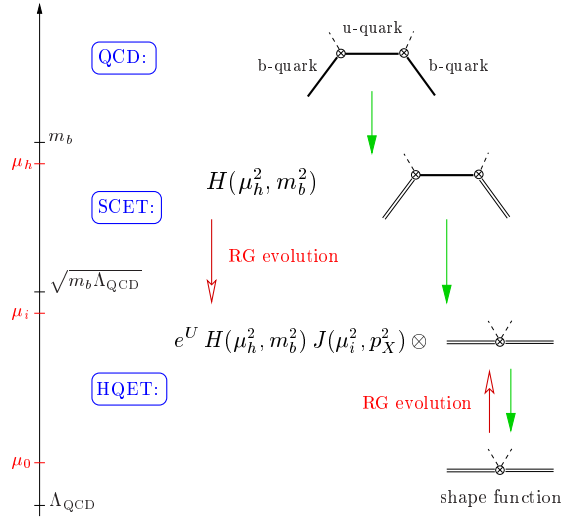, width=7.4cm}
\vspace{-5mm}
\caption{\label{fig:factmeth} Contributions from different energy
scales are factorized by matching onto Effective Field Theories: 
QCD $\to$ SCET $\to$ HQET.}
\end{figure}

Recently, a systematic framework has been developed
\cite{Bauer:2003pi,Bosch:2004th,Bosch:2004bt} that enables us to
compute the differential decay rates in the shape-function region. A
series of matching calculations QCD $\to$ SCET $\to$ HQET is necessary
to disentangle physics at the three different energy scales $m_b$,
$\sqrt{m_b\Lambda_{\rm QCD}}$, and $\Lambda_{\rm QCD}$. The general
methodology is schematically visualized in Figure~\ref{fig:factmeth}.

The idea is as follows: Under the assumption of quark-hadron duality,
the inclusive decay rate can be calculated in QCD using the optical
theorem. The intermediate line denotes the energetic $u$-quark
propagator. At a hard scale $\mu_h \sim m_b$, the amplitude is matched
onto SCET, which correctly describes the infra-red degrees of freedom
below this scale. Physics effects from scales above $\mu_h$ are
contained in a hard coefficient function $H(\mu_h,m_b)$. Since the
$u$-quark momentum is parametrically off-shell by an amount of 
$O(m_b \Lambda_{\rm QCD})$, the corresponding propagator can be
integrated out at an intermediate matching scale
$\mu_i\sim\sqrt{m_b\Lambda_{\rm QCD}}$, leading to a perturbatively
calculable jet function $J$ and a low-energy description in HQET. The
matrix element of the resulting leading power operator defines the
hadronic shape function, which cannot be computed using analytic
techniques. Large (Sudakov) logarithms are resummed when evolving the
hard function from the hard scale $\mu_h$ down to the intermediate
scale $\mu_i$. Finally the decay amplitude is expressed as a
convolution integral of perturbatively calculable functions and the
shape function renormalized at $\mu_i$.

Event fractions and spectra are derived by performing the necessary
phase-space integrations, {\em before} integrating over the shape
function. In this way, the results are model independent and once
again given as convolution integrals over the shape function.

\section{PROPERTIES OF THE SHAPE FUNCTION}

The shape function is a non-perturbative structure function that
encodes the Fermi motion of the heavy quark inside the $B$ meson
\cite{Neubert:1993ch}. Although the functional form of it cannot be
derived using analytical techniques, it is nevertheless possible to study
other properties of it using perturbation theory. In particular, the
dependence on the renormalization scale $\mu$ can be reliably computed
as long as $\mu$ is much larger than $\Lambda_{\rm QCD}$. At leading
order in renormalization-group improved perturbation theory it is
given by the simple formula (for $\mu_i\ge\mu_0$)
\begin{equation}\label{wow}
   \hat S(\hat\omega,\mu_i) = \frac{e^{V_S(\mu_i,\mu_0)}}{\Gamma(\eta)} 
   \int_0^{\hat\omega}\!d\hat\omega'\, \frac{\hat S(\hat\omega',\mu_0)}
       {\mu_0^{\eta}\,(\hat\omega-\hat\omega')^{1-\eta}} \,,
\end{equation}
where $\eta = (16/25)\ln(\alpha_s(\mu_0)/\alpha_s(\mu_i))$, and the
exponent $V_S$ vanishes in the limit $\mu_i\to\mu_0$. 
(The hatted notation denotes that the shape function is defined in a
scheme-independent way and has support for 
$\hat\omega \in [0,\infty[$. For details, see Ref.~\cite{Bosch:2004th}.)  
As a consequence of (\ref{wow}), radiative corrections build up a tail
that vanishes slower than $\hat\omega^{-1}$ for
$\hat\omega\to\infty$. This means that moments of the shape function
(including its norm) are UV-divergent. This is, however, not an
obstacle in practice, where the shape function is only needed over a
finite interval. It is then natural to define moments accordingly over
a finite integration domain, by means of a cutoff. Expanding them in a
local OPE allows us to extract information about the shape function
from the measurement of HQET parameters such as $\bar\Lambda$ and
$\mu_\pi^2$. The shape-function mass scheme \cite{Bosch:2004th} is
convenient for such calculations and free of renormalon
ambiguities. We stress, however, that information of $\bar\Lambda$ and
$\mu_\pi^2$ in {\em any} low-scale subtracted mass scheme is useful,
since there exist perturbatively calculable relations between these
mass definitions.

A somewhat surprising observation is made by inspecting the behaviour
of the moments under variation of the cutoff. It turns out that their
values {\em decrease} for larger cutoffs. It follows that the tail of
the shape function must be negative! In fact, it is possible to derive
the explicit form of the shape function itself for large values of
$\hat\omega\gg\Lambda_{\rm QCD}$. We find that the asymptotic tail is
given in the $\overline{\rm MS}$ scheme as \cite{Bosch:2004th}
\begin{eqnarray}
\hat S(\hat\omega,\mu) 
  & \stackrel{\hat\omega\gg\Lambda_{\rm QCD}}{\longrightarrow} &
  -\frac{C_F\alpha_s(\mu)}{\pi}\,\frac{1}{\hat\omega-\bar\Lambda} 
  \times \nonumber \\
  && \left(2\ln \frac{\hat\omega-\bar\Lambda}{\mu}+1\right)\,
  +\ldots\,,
\end{eqnarray}
and one has to drop the common interpretation of the renormalized shape
function as a probability distribution. 

\subsection{MODEL SHAPE FUNCTIONS} \label{sec:modelSF}

\begin{figure}[b!] 
\epsfig{file=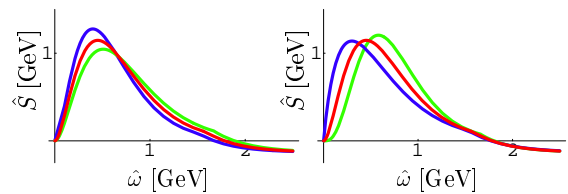, width=7.4cm}
\vspace{-5mm}
\caption{\label{fig:SFmodel} Model shape function under variation of
the numerical values for $\bar\Lambda$ and $\mu_\pi^2$ within their
errors. Correlated variations are shown on the left, anti-correlated
on the right.}
\end{figure}

In the region $\hat\omega \sim \Lambda_{\rm QCD}$, where it is
relevant for phenomenological applications, the shape function cannot
be computed within perturbation theory. In the following we will use a
simple model that is consistent with all analytic constraints
mentioned above.  The model is constructed in such a way that its
first few moments are consistent with the experimental values
$\bar\Lambda = (0.63\pm 0.07)\; \rm GeV$ and 
$\mu_\pi^2 = (0.27\pm 0.07)\; \rm GeV^2$ at a reference scale of
$\mu_i = 1.5\;\rm GeV$ in the shape-function scheme (see
Ref.~\cite{Bosch:2004th} and references therein). Below we will use
nine different functions (corresponding to the nine different pairs of
values $(\bar\Lambda,\mu_\pi^2)$ when varied within their errors) to
account for our ignorance of the functional form of the shape
function. Figure~\ref{fig:SFmodel} is intended to give the reader an
idea of such variations. We stress, however, that a hypothetical
knowledge of the first few moments to arbitrary precision would not
determine the functional form of the shape function.

\section{EVENT FRACTIONS WITH CUTS ON $P_+$ OR $M_X$}

With the methodology outlined in Section~\ref{sec:method} we find the
following expression for the fraction of events that survive a cut
$P_+ \le \Delta_P$, i.~e.~that are located below a horizontal line in
the phase-space picture in Figure~\ref{fig:phasespace}. At leading
power in $\Lambda_{\rm QCD}/M_B$ and at next-to-leading order in
renormalization-group improved perturbation theory, it is given as a
weighted integral over the shape function:
\begin{eqnarray}
   F_P(\Delta_P)&=&T(a)\; e^{V_H(\mu_h,\mu_i)} \;\times \nonumber \\
&& \int\limits_0^{\Delta_P}d\hat\omega\,\hat S(\hat\omega,\mu_i) 
   W_P(\hat\omega,\Delta_P,a)\;.
\end{eqnarray}
Large Sudakov logarithms are resummed in the expression 
$T(a)\; e^{V_H(\mu_h,\mu_i)}$ (where $a=16/25 \times
\ln(\alpha_s(\mu_i)/\alpha_s(\mu_h)) \approx 0.3$ for typical choices
of $\mu_h,\mu_i$), and the weight function $W_P$ is perturbatively
calculable \cite{Bosch:2004th} and starts as $W_P = 1+O(\alpha_s)$.

When cutting on the hadronic invariant mass $M_X^2 \le s_0$ instead, the
corresponding event fraction is given by
\begin{eqnarray}\label{eq:FM}
   F_M(s_0)&=& F_P(s_0/M_B)+ T(a)\; e^{V_H(\mu_h,\mu_i)} 
   \;\times \nonumber \\
&& \int\limits^{\sqrt{s_0}} d\hat\omega\,\hat S(\hat\omega,\mu_i) 
   W_{\rm triangle}(\hat\omega)\;.
\end{eqnarray}
The two terms in the above expression can be easily interpreted using
the phase-space picture in Figure~\ref{fig:phasespace}. For an optimal
cut $s_0 = M_D^2$ the allowed region (dark gray) splits into two
distinct areas, separated by the dashed line. Events below that line
are included in the count by the first term in (\ref{eq:FM}). The
second term counts events in the triangle-shaped region above the
dashed line. To account for the tip of this triangle
($P_+=P_-=\sqrt{s_0}$) is problematic in the aforementioned framework,
because the assumption $P_+ \ll P_-$ and the resulting power-expansion
rules are no longer valid. However, only few events are located near
the tip, which is reflected in the finding that the corresponding
contribution to $F_M$ scales like $(\Lambda_{\rm QCD}/M_B)^{(3-a)/2}$
(assuming $s_0\sim \Lambda_{\rm QCD} M_B$). Therefore one obtains a
well defined leading power expression by setting the upper limit of
integration from $\sqrt{s_0}$ to $\infty$ \cite{Bosch:2004th}. It is
unclear, however, how to treat power corrections to $F_M$
systematically in the future improvement of this result.

\begin{figure*}[t!]
\begin{center}
\epsfig{file=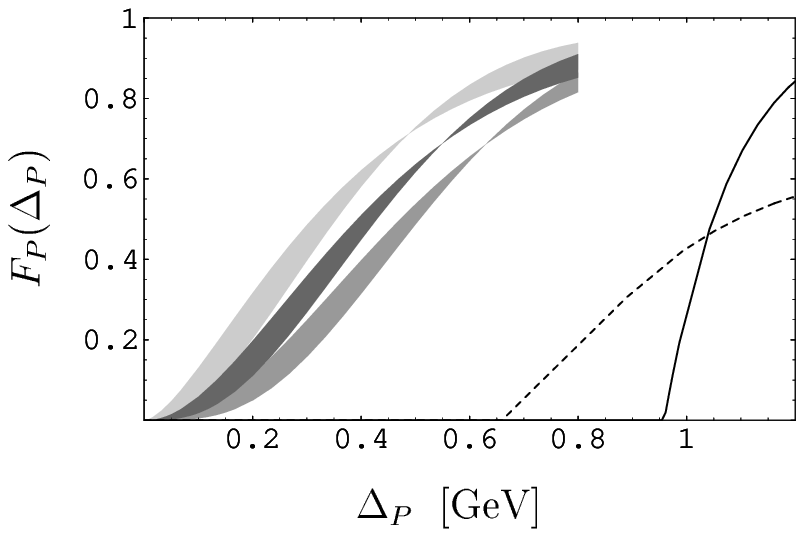, width=7cm}
\hfill
\epsfig{file=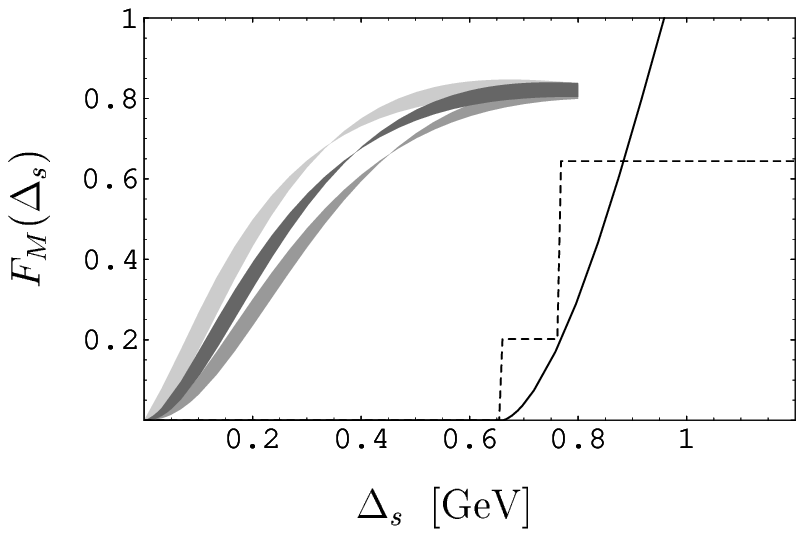, width=7cm}
\vspace{-5mm}
\caption{\label{fig:PplusAndMX} Model Predictions for $F_P(\Delta_P)$
and $F_M(\Delta_s)$ with $\Delta_s = s_0/M_B$. Each gray band
corresponds to a fixed value of $\bar\Lambda(\mu_i,\mu_i)$, while the
value of $\mu_\pi^2(\mu_i,\mu_i)$ is varied within its error bars. The
solid lines on the right in each graph denote the inclusive $B\to X_c$
background, while the dashed lines indicate exclusive $B\to D^{(*)}$
decays.}
\end{center}
\end{figure*}

\subsection{MODEL-INDEPENDENT RELATIONS}

QCD-Factorization is a powerful tool because the non-perturbative
structure functions entering the theoretical description are
universal, process-independent quantities. The shape function needed
in this discussion on $B\to X_u\,l^-\bar\nu$ decays is also the only
leading-power structure function entering the calculation of the
important $B\to X_s\,\gamma$ decay rate. In fact, at leading
power and leading order in $\alpha_s$, the shape function is identical
to the photon spectrum. This implies that it is possible to construct
relations between the photon spectrum and the event distribution
functions in $B\to X_u\,l^-\bar\nu$ decays in which the shape function
has been eliminated. An example of such a relation is
\cite{Bosch:2004bt}
\begin{equation}
F_P(\Delta) = \int\limits^{\frac{M_B}{2}}_{\frac{M_B-\Delta}{2}} 
  dE_\gamma \, \frac{1}{\Gamma_s} \frac{d\Gamma_s}{dE_\gamma} \; 
  \underbrace{w(\Delta,E_\gamma)}_{1 + \alpha_s \ln \ldots}\;.
\end{equation}
The weight function $w$ is perturbatively calculable once our
formalism has been applied to calculate the $B\to X_s\,\gamma$ photon
spectrum. Note that a similar relation to the hadronic invariant mass
distribution $F_M$ would be possible, but more complicated. Because of
the contribution from the triangle region, the photon spectrum would
be needed over a larger window and beyond the region where it is
experimentally known with acceptable precision.

\subsection{MODEL PREDICTIONS}

We use the nine model shape functions discussed in
Section~\ref{sec:modelSF} to predict the event fractions $F_P$ and
$F_M$ and demonstrate their sensitivity to the shape function. In
principle, one also has to vary the functional form of the shape
function without violating the constraints for its first few
moments. In the end we find, however, that such variations are already
covered in the sample of the nine models \cite{Bosch:2004th}.

In Figure~\ref{fig:PplusAndMX} our results are depicted as gray
bands. Each of the bands represents the results using shape function
models that correspond to a fixed value of $\bar\Lambda$, while the
numerical value of $\mu_\pi^2$ is varied within its error bars. We
draw three bands for three values of $\bar\Lambda$: 0.56 GeV, 0.63
GeV, and 0.70 GeV, in accordance with our discussion in
Section~\ref{sec:modelSF}. The calculations are valid for $\Delta_P$
and $\Delta_s = s_0/M_B$ of order $\Lambda_{\rm QCD}$. If these values
are too small, i.~e.~parametrically of order $\Lambda_{\rm QCD}^2/M_B
\approx 50\;\rm MeV$, the assumption of quark-hadron duality is no
longer justified. If, on the other hand, $\Delta_{P,s}$ are larger
than about 0.8 GeV, the collinear expansion used in the
calculation breaks down.

\begin{table}[b!]  
\begin{center}
\caption{\label{tab:PpAndMX}Shape-function uncertainties for typical
cuts on the hadronic variables $M_X$ and $P_+$.}
\begin{tabular}{ll}
\hline
Cut & Efficiency \\
\hline
$M_X\le M_D$ & $({81.4_{\,-3.7}^{\,+3.2}})\%$ \\
$M_X\le(1.7\,\mbox{GeV})$ & $(78.2_{\,-5.2}^{\,+4.9})\%$ \\
$M_X\le(1.55\,\mbox{GeV})$ & $(72.7_{\,-6.3}^{\,+6.4})\%$ \\
\hline
$P_+\le\frac{M_D^2}{M_B}=0.66$\,GeV & $({79.6_{\,-8.2}^{\,+8.2}})\%$ \\
$P_+\le 0.55$\,GeV & $(69.0_{\,-12.1}^{\,+9.7})\%$ \\
\hline
\end{tabular}
\end{center}
\end{table}

Our findings are summarized in Table~\ref{tab:PpAndMX} below. For
``ideal'' cuts, i.~e.~where the charm background starts, 
$\Delta_{P,s} = M_D^2/M_B \approx 660\;\rm MeV$. The results show that
cutting on the hadronic invariant mass $M_X\le M_D$ is rather
insensitive to shape-function effects. However, due to detector
resolution effects, it is typically necessary to lower the cut and
move away from the point where the charm background starts. In that
case, the uncertainties due to our ignorance of the shape function
become quickly larger, and are of comparable size to the uncertainties
in an ``ideal'' $P_+$ cut. It seems a clear advantage for cutting on
hadronic invariant mass. However, we will argue below that it might
not be necessary to move too far away from the charm background when
cutting on $P_+$. In that case the hadronic invariant mass cut does
not offer that advantage anymore and both methods give high
efficiencies of about 70-80\% with a relative uncertainty of 
roughly 10\%.

\section{CUTTING ON THE CHARGED LEPTON ENERGY}

\begin{figure}[t!]
\epsfig{file=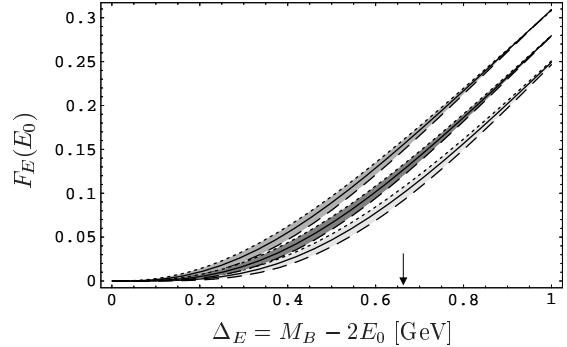, width=7.4cm}
\vspace{-5mm}
\caption{\label{fig:El} Model Predictions for the event fraction
$F_E$. The bands are explained in Figure~\ref{fig:PplusAndMX}, and the
arrow indicates the start of the charm background.}
\end{figure}

A prominent alternative way to discriminate the charm background is to
cut on the charged-lepton energy. This is experimentally favored
because it does not require the reconstruction of the neutrino
momentum. (Both cutting on $P_+$ or $M_X$ need a neutrino
reconstruction.) Unfortunately, the method is not favored by
theorists, for several reasons. The most obvious one is that far fewer
events survive such a cut. In our prediction this is reflected by the
fact that the weight function is suppressed by a power of
$\Lambda_{\rm QCD}/M_B$:
\begin{eqnarray}
F_E(\Delta_E)&=&T(a)\; e^{V_H(\mu_h,\mu_i)} \;\times \nonumber \\
&& \hspace{-15mm} \int\limits_0^{\Delta_E}d\hat\omega\,
   \hat S(\hat\omega,\mu_i)\,
   \frac{2(\Delta_E-\hat\omega)}{M_B-\hat\omega}\;
   [1+O(\alpha_s)]\;,
\end{eqnarray}
where $\Delta_E = M_B-2E_0$ and $E_0$ is the lower limit of allowed
lepton energy. Because of the rather low efficiency, this method is
more prone to uncertainties from other effects, e.g.~weak annihilation
\cite{Voloshin:2001xi}. \linebreak
\begin{table}[b!]
\begin{center}
\caption{\label{tab:El} Predictions and shape-function uncertainties
for a cut on the charged lepton energy. For comparison, the results
using the DeFazio/Neubert (DFN) model \cite{DeFazio:1999sv} are also
given.}
\begin{tabular}{llll} \hline 
cut $E_0$ & $\Delta_E$ & DFN  & BLNP  \\
$$ [GeV] & [GeV] & [\%] & [\%] \\ \hline 
$E_l \ge 2.31$ & 0.66 & \phantom{0}7.9$^{+3.4}_{ -2.2}$ & 
12.5$^{+3.4}_{ -3.5}$ \\
$E_l \ge 2.2$  & 0.88 & 14.4$^{+4.4}_{ -3.3}$ & 
22.2$^{+3.2}_{ -3.6}$ \\
$E_l \ge 2.1$  & 1.08 & 20.9$^{+4.9}_{ -4.1}$ & 
31.7$^{+3.0}_{ -3.1}$ \\ \hline 
\end{tabular}
\end{center}
\end{table}
\indent Numerical predictions using shape-function models are
visualized in Figure~\ref{fig:El} and summarized in
Table~\ref{tab:El}. For orientation, the beginning of the charm
background at $\Delta_E = 660\;\rm MeV$ is marked with an arrow in the
Figure. Evidently, the relative shape-function uncertainties are
large, and do not show a ``focusing effect'' as in the case of $P_+$
or $M_X$ cuts. Because of the large shape-function sensitivity and the
overall small efficiency, the extraction of $|V_{ub}|$ from the
charged-lepton energy endpoint region is theoretically disfavored.

An interesting observation can be made when comparing the third column
in Table~\ref{tab:El} with our results (BLNP = Bosch, Lange, Neubert,
Paz \cite{Bosch:2004th}) in the fourth column. The former are
predictions using the familiar DeFazio/Neubert model (DFN) of
Ref.~\cite{DeFazio:1999sv}, where a simple replacement rule is used to
implement the shape function into the results of parton-level
calculations. The crucial difference is that our framework allows for
a clean separation of short- and long-distance physics effects. In
order to demonstrate this, consider the one-loop order result of the
triple differential decay rate
\cite{Bauer:2003pi,Bosch:2004th,DeFazio:1999sv}. The leading logarithm
comes with a coefficient of $-4$. In the framework of factorization,
this coefficient will split into $-4=-8+4$. While the leading
logarithm of the jet function comes with the coefficient $+4$, the
$-8$ will be absorbed in the ``partonic'' shape function. The simple
replacement rule of the DFN model does not capture this effect
correctly. In addition to this important observation, the
factorization framework allows for a systematic Sudakov resummation,
which was not performed in the work of \cite{DeFazio:1999sv}.

The numerical impact of these improvements over our previous
understanding is that the predictions for the event fractions in the
charged-lepton energy endpoint have increased quite noticeably. This
implies in turn that experimental $|V_{ub}|$ determinations from the
charged-lepton energy endpoint region might need to be corrected to
lower values. In general, our analysis for $B\to X_u\,l^-\bar\nu$
decay spectra in the shape-function region suggests that more events
than previously anticipated are located in the shape-function
region. It will be most interesting to study how power corrections
will affect this conclusion.

\section{CHARM BACKGROUND}

\begin{figure}[!ht]
\epsfig{file=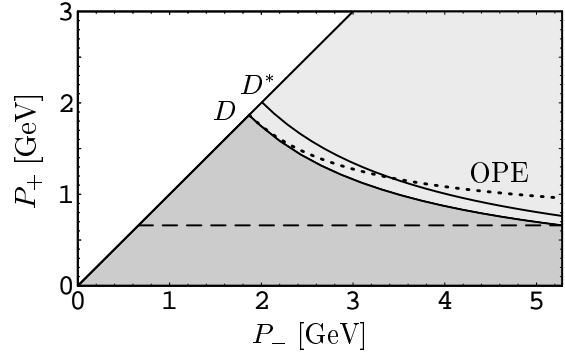, width=7.4cm}
\vspace{-5mm}
\caption{\label{fig:charmBG} Phase-space picture for the charm
background. The exclusive $B\to D$ events are located on the lower
solid line. The OPE prediction for inclusive $B\to X_c$ events are
located above the dashed line. The gap is filled with exclusive events
$B\to D^*$, $B\to D\pi$, etc. }
\end{figure}

The $P_+$ spectrum has a considerable advantage over the hadronic-mass
spectrum when considering the charm background. It is straight-forward
to visualize the area in phase space that is populated by inclusive
$B\to X_c$ decays. The OPE prediction uses quark-hadron duality and
the fact that $p_+ p_- \ge m_c^2$ in the parton picture. Here, 
$p_\pm = P_\pm - \bar\Lambda$ are the light-cone components of the
parton momentum, and $m_c$ is the charm quark mass. The dotted line in
Figure~\ref{fig:charmBG} marks the threshold $p_+ p_- = p^2=m_c^2$. At
tree-level, all events are located on that line. Radiative corrections
smear the event distribution into the area above it. In this sense,
the tree-level scenario serves as an upper bound of the inclusive
charm background.  Note that the dotted line touches the tip of the
triangle, $P_+=P_-=M_D$ and extends to the right while always staying
{\em above} the exclusive $B\to D$ solid line. It follows that the
variable $P_+$ is bound to be always greater than approximately 960
MeV in the parton model. Therefore an ideal cut on $P_+ = 660$ MeV has
the advantage of a comfortable ``buffer zone'' to the inclusive
background.

The gap between the charm-free area (dark gray) and the inclusive
charm background (above the dotted line) is filled with exclusive
final states such as $B\to D$, $B\to D^*$, $B\to D\pi$, etc. 

In Figure~\ref{fig:PplusAndMX} the results for a tree-level
calculation \cite{Bosch:2004bt} of the inclusive and exclusive event
fractions are drawn as solid and short-dashed lines,
respectively. Note that the event fractions are normalized to unity;
however, the reader should keep in mind that the $B\to X_c$ background
is about 60 times larger than the $B\to X_u$ signal. Consider now what
happens when the quantities $\Delta_{P,s}$ approach the charm
threshold near 660 MeV. The charmed hadronic invariant mass spectrum
receives step increments when the $D$ and $D^*$ phase space becomes
available. At the same time, the inclusive rates start near 660 MeV
and reaches unity at 960 MeV in the tree approximation. Detector
resolution effects smear the charm events into the region below 660
MeV, and force us to move away from the ideal cut, which leads to
enhanced shape-function uncertainties in the $B\to X_u$ event
fractions (see Table~\ref{tab:PpAndMX}).

The scenario for the $P_+$ spectrum is different. The short-dashed
line in the left part of Figure~\ref{fig:PplusAndMX} also starts near
660 MeV, but has a smooth onset. The same applies when the $D^*$ phase
space becomes available, leading to a (tiny) kink in the
line. However, up until this point, the charm background consists 
{\em exclusively} of $B\to D$ events, since the inclusive events start
much later at 960 MeV. The knowledge of the precise nature of the
background ought to help experimentally to stay closer to the ideal
cut, because it can be modeled with greater precision. We believe that
this fact, together with the cleanliness of the theoretical
description, make the $P_+$ method an excellent high-efficiency
candidate for a determination of $|V_{ub}|$, and we urge the
experimental community to perform a measurement of the $P_+$ spectrum.

\section{CONCLUSION}

In this talk we have presented recent advances in the understanding of
inclusive $B$ decays in the shape-function region.  Using effective
field theory techniques, a systematic framework was developed, where a
simple OPE cannot be used due to soft shape function effects. New
results for event distributions of several important kinematic
quantities, including the hadronic invariant mass and the
charged-lepton energy, have been presented. We advertised a new method
for the extraction of $|V_{ub}|$ based on the $P_+$ spectrum. We
stress, however, that this should not be understood as a
discouragement of pursuing other methods, but rather as an
encouragement to consider the $P_+$ method as a powerful alternative.

The calculations have been performed in the heavy-quark limit and at
next-to-leading order in RG-improved perturbation theory. Our
prediction for the fraction of events with the optimal cut 
$P_{+}\le M_{D}^{2}/M_{B}$ is \cite{Bosch:2004bt}
\begin{equation}\label{FPnum}
   F_{P} = (79.6\pm 10.8\pm 6.2\pm 8.0)\% \,,
\end{equation}
where the errors represent the sensitivity to the shape function, an
estimate of ${\cal O}(\alpha_{s}^{2})$ contributions, and power
corrections, respectively. Each of these uncertainties
can be reduced in the future within the general framework presented in
\cite{Bauer:2003pi,Bosch:2004th,Bosch:2004bt} and outlined in
Section~\ref{sec:method}.  Much can be learned about the shape
function from the $B\to X_s \gamma$ photon spectrum, once the
formalism has been applied. The calculation of 
${\cal O}(\alpha_{s}^{2})$ corrections would be a major effort, but is
feasible in the future. The important question of power corrections
are currently investigated by several groups in the theory community.

The CKM-matrix element $|V_{ub}|$ can be extracted by comparing a
measurement of the partial rate $\Gamma_{u}(P_{+}\le\Delta_{P})$ with
a theoretical prediction for the product of the event fraction
$F_{P}(\Delta_P)$ and the total inclusive $\bar B\to X_{u} \,
l^{-}\bar\nu$ rate. The resulting theoretical uncertainty on
$|V_{ub}|$ is
\begin{equation}
   \frac{\delta |V_{ub}|}{|V_{ub}|} = (\pm 7\pm 4\pm 5\pm 4)\% \,,
\end{equation}
where the last error comes from the uncertainty in the total rate
\cite{Hoang:1998ng,Uraltsev:1999rr}. Because of the large efficiency
of the $P_{+}$ cut, weak annihilation effects \cite{Voloshin:2001xi}
have an influence on $|V_{ub}|$ of less than 2\% and can be safely
neglected.

\section*{ACKNOWLEDGMENTS}

It is a pleasure to acknowledge my collaborators Stefan Bosch,
Matthias Neubert, and Gil Paz. I would also like to thank Jon Rosner
for inviting me to give this talk, and the organizers of BEACH 2004
for a beautiful conference. This research was supported by the
National Science Foundation under Grant PHY-0098631.

\end{document}